\documentclass[%
preprint,
superscriptaddress,
amsmath,amssymb,
aps,
pre,
]{revtex4-2}
\usepackage[dvipdfmx]{graphicx}
\usepackage{bm}
\usepackage{amsmath}
\usepackage{amsfonts}
\usepackage{amssymb}
\usepackage{comment}
\usepackage{color}

\begin{document}
\title{Extensive tip-splitting of injected organic liquid into an aqueous viscoelastic fluid} 

\author{Kiwamu Yoshii}
\affiliation{%
	Department of Applied Physics, Tokyo University of Science, 6-3-1 Nijuku, Katsushika-ku, Tokyo, 125-8383, Japan}
\affiliation{
	Earthquake Research Institute, The University of Tokyo, 1-1-1 Yayoi, Bunkyo-ku, Tokyo 113-0032, Japan
} 
\affiliation{Department of Physics, Nagoya University, Furo-cho, Chikusa, Nagoya, Aichi 464-8602, Japan}

\author{Kojiro Otoguro}%
\affiliation{%
	Department of Applied Physics, Tokyo University of Science, 6-3-1 Nijuku, Katsushika-ku, Tokyo, 125-8383, Japan}
	
\author{Ayane Pygoscelis Sato}%
\affiliation{%
	Department of Applied Physics, Tokyo University of Science, 6-3-1 Nijuku, Katsushika-ku, Tokyo, 125-8383, Japan}

\author{Yutaka Sumino}%
\email{ysumino@rs.tus.ac.jp}
\affiliation{%
	Department of Applied Physics, Tokyo University of Science, 6-3-1 Nijuku, Katsushika-ku, Tokyo, 125-8383, Japan}
\affiliation{
	Water Frontier Science \& Technology Research Center and I$^2$ Plus, Research Institute for Science \& Technology, Tokyo University of Science, 6-3-1 Nijuku, Katsushika-ku, Tokyo, 125-8585, Japan
}

\date{\today}
\begin{abstract}
The injection of a fluid into another fluid causes a spatiotemporal pattern along the injection front. {\color{black} Viscous fingering is a well-known example when the replaced material is a viscous fluid.  Notably, most fluids are, in reality, viscoelastic, i.e., they behave as an elastic solid over short timescales. For this reason, it is important to study the situation when the replaced fluid is viscoelastic.} In this study, we observe extensive tip-splitting in the fingering pattern when an incompressible organic liquid was injected into an oleophilic Hele--Shaw cell filled with an aqueous viscoelastic fluid made of a wormlike micellar solution. The tip-splitting led to thin fingers with a characteristic size comparable to four times the cell thickness. We examined the material properties and suggest that the thin fingering pattern observed in our current system is due to the delamination of viscoelastic fluid from the bottom substrate surface. Our result shows that the effect of interfacial energy in the existing solid layer should be considered in the injection process.
\end{abstract}

\maketitle

\section{\label{sec:level1}Introduction}
When a fluid is injected into a confined space filled with another fluid, existing fluids are displaced and often create complex pattern dynamics. Here, the injection of fluid introduces the non-equilibrium condition through kinetic energy, momentum, and chemical components, leading to the appearance of a dissipative structure created under far from equilibrium conditions~\cite{Nicolis1977}.  
The injection of fluid into a thin cell, known as a Hele--Shaw cell, filled with another viscous fluid exhibits viscous fingering~\cite{Homsy1987, DeWit2005}. Recent studies on viscous fingering have adopted the use of chemical reactions~\cite{Haudin2014, Wagatsuma2017, Shukla2016, Tanaka2023a}, or miscibility of the fluid~\cite{Mishra2009, Chui2015}.

{\color{black} Viscous liquids are useful yet only provide an approximation. Actual fluids can have elastic aspects, which we call viscoelasticity.} Understanding the impact of viscoelasticity on macroscopic dynamics is not only theoretically interesting but also important for industries like cosmetics and food manufacturing. The same goes for geophysical studies. Recent research on earthquakes, as well as slow earthquakes~\cite{Obara2016}, has revealed that the migration of water (a viscous liquid) in a subducting mantle (a viscoelastic material) can trigger earthquakes~\cite{Ujiie2018, Suzuki2014, Tanaka2018}. Therefore, it is crucial to study the effect of viscoelasticity to get a better understanding of these phenomena.

\begin{figure}
\centering
\includegraphics{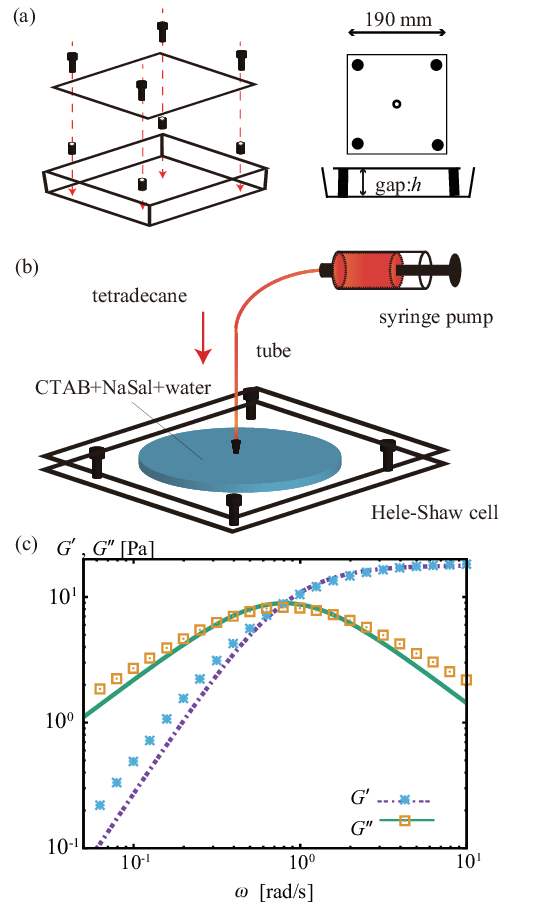}
\caption{\label{fig:1}(a)~Schematic representation of a Hele--Shaw cell made from a square Petri dish. The gap width between the two parallel plates was $h=$ 0.5 mm. (b) Schematics of the experimental system. A Hele--Shaw cell was filled with the viscoelastic outer fluid, an aqueous mixture of CTAB and NaSal, and stored for {\color{black} 3} h. The inner fluid, tetradecane stained with oil red, was then injected steadily from the center of the cell. (c) Storage modulus $G'$ and loss modulus $G''$ of the viscoelastic outer fluid. The lines represent a fit by a Maxwell model with a single relaxation time, which gives us the shear modulus {\color{black} $G = 17.85 \pm 0.17$~Pa and the relaxation time $\tau = 1.23 \pm 0.03$~s}.
}
\end{figure}

For the above situation, it is relevant to investigate systems in which a viscous fluid is injected into a viscoelastic fluid. In this context, the injection of a fluid into an agarose gel has been studied~\cite{Hirata1998}. Similarly, recent studies on viscoelastic fluids have also been conducted~\cite{Zhao1993, Saintyves2019, Hue2022}. {\color{black} In their study, the main focus was given to the transition from viscous to elastic behavior.  For this reason, the injected fluid in these studies often had similar wetting properties to the displaced fluid. In this study, we focus on the elastic outer fluid with special attention on the wetting of the substrate surfaces.} Thus, we inject an organic liquid into a cell made of oleophilic substrates filled with aqueous viscoelastic fluid of a wormlike micellar solution. This solution is composed of a mixture of cetyltrimethylammonium bromide (CTAB) and sodium salicylate (NaSal) and is highly viscous~\cite{Rodrigues2008, AnhTuan2013} and also elastic~\cite{Shikata1987, Shikata1988, Gladden2007,Sumino2012a, recipe}, even with only a few percent of additional chemicals to water. 

In this paper, we describe the results from an experiment under the geometry of Hele--Shaw cells. An incompressible organic liquid is injected into a Hele--Shaw cell filled with a wormlike micellar solution. {\color{black} The injection speed is varied as a parameter but kept high enough to have the wormlike micellar solution elastic, and the spatiotemporal dynamics of the injection patterns are observed.} The fingering pattern showed thick and thin fingers, and thin fingers are characterized by extensive tip-splitting. Based on the observed dynamics, the rheological relaxation time, and the wetting behavior, we suggest that thin fingering may represent the delamination of the viscoelastic fluid from the substrate. 
{\color{black} Interestingly, the wetting of elastic fluid on a substrate surface, which is typically the character of a liquid, plays an essential role in the fingering process.} 

\begin{figure*}[b]
\centering
\includegraphics[width=150mm]{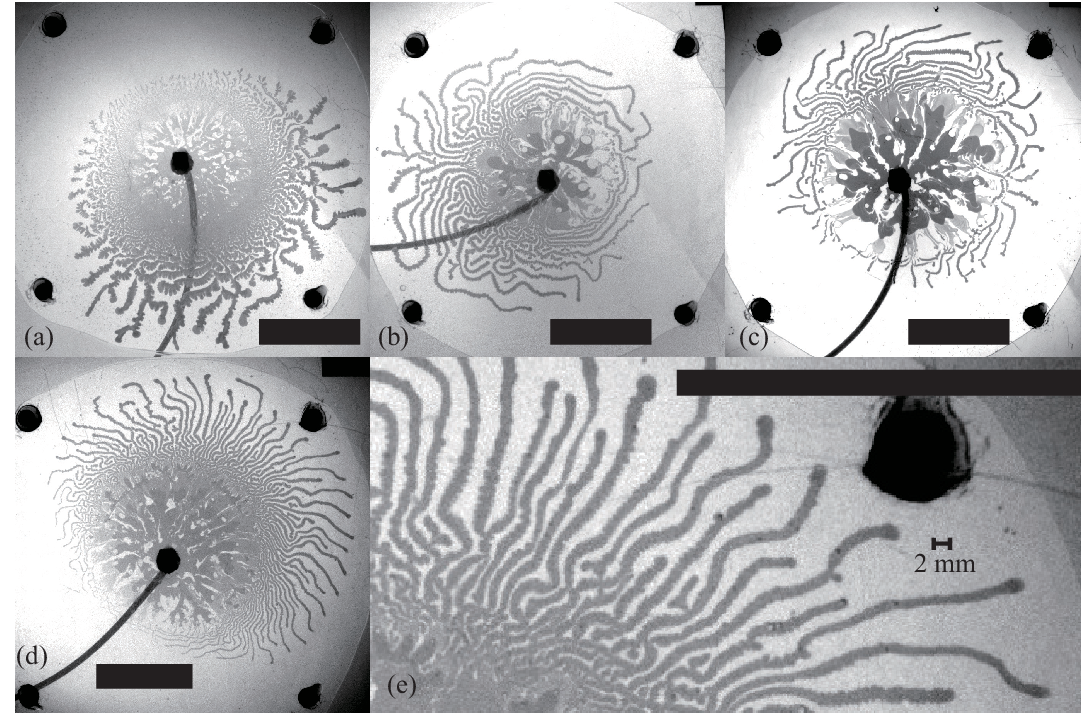}
\caption{\label{fig:2}Snapshots of typical fingering patterns, where {\color{black} $Q=$ (a) 1.0, (b) 3.0, (c) 6.0, and (d, e) 9.0 mL/min. Scale bar: 50 mm. The image in (e) is an enlarged image of (d). The snapshots were taken just before the inner fluid reached the edge of the outer fluid, which was (a) 60.00 s, (b) 20.00 s, (c) 10.00 s, and (d, e) 6.67 s after the start of the injection. }
We observed thin fingers characterized by extensive tip-splitting for all conditions, as well as thick fingers observed near the inlet.}
\end{figure*}

\section{\label{sec:level2}Experimental system}
The experimental system was a thin horizontal cell (Hele--Shaw cell) with a gap width of $h=$~0.5 mm. The cell was made using a polystyrene Petri dish (166058, Thermo Fisher Scientific) as the lower plate and an acrylic plate (Mitsubishi Rayon Co., Ltd.) as the upper plate~(Fig.~\ref{fig:1}(a)). A Luer fitting (ISIS Co., Ltd.) was fixed at the center of the upper plate and used as an inlet for injecting inner fluids. 

We purchased cetyltrimethylammonium bromide (CTAB) from Tokyo Chemical Industry Co., Ltd. and sodium salicylate (NaSal), tetradecane, and oil red from Wako Pure Chemical Industries Ltd. Fluorocarbon surface modifier was obtained from Fluoro technology (Fluorosurf; FS-1090J-2.0). {\color{black} We purified CTAB with the recrystallization method three times using a mixture of acetone and methanol.}
{\color{black} The inner fluid consisted of tetradecane colored with 0.2 wt.\% oil red.}
The outer fluid consisted of {\color{black} 1.82 wt.\% CTAB and 0.80 wt.\% NaSal}, dissolved in pure water purified using the Millipore Milli-Q system. The CTAB and NaSal were mixed into the water while the temperature was 60 $^{\circ}$C. 

The Hele--Shaw cell was filled with approximately 10 mL of the outer fluid and stored for {\color{black} 3 h} before the experiment. Then, the inner fluid was injected from the center of the upper plate (Fig.~\ref{fig:1}(b)). A syringe pump (CXF1010; ISIS Co., Ltd.) was used to inject the inner fluid at a nominally constant injection rate $Q$. We used a glass syringe (Tsubasa Industry Co., Ltd.) and a nylon tube (Nihon Pisco Co., Ltd.; diameter 2.5 mm) to ensure the high rigidity of the injection environment.
$Q$ was varied from 1.0--9.0 mL/min. The pattern dynamics were recorded from the bottom of the cell using a digital video camera {\color{black} (DMK37BUX273, The Imaging Source) at 16 bit and 30 Hz} and analyzed using the Image J software~\cite{imagej}.

The outer fluid is known to form a wormlike micellar solution and behave as a Maxwellian viscoelastic fluid with a single relaxation time~\cite{Shikata1987, Shikata1988, Gladden2007, hU1994, Inoue2005}. The rheology of the outer fluid was measured from its oscillatory shear using a rheometer {\color{black}(MCR-302, Anton Paar) with cone plate (CP25-2, Anton Paar). We conducted linear oscillatory measurement with the amplitude of 10\% shear at 20 $^\circ$C}. The storage modulus $G^{\prime}$ and loss modulus $G^{\prime\prime}$ are shown in Fig.~\ref{fig:1} (c). {\color{black} We estimated the shear modulus $G$ and the relaxation time $\tau$ by fitting these observed data to the equation $G^{\prime}(\omega)=(G\omega^2\tau^2)/(1+\omega^2\tau^2)$ and~$G^{\prime\prime}(\omega)=(G\omega\tau)/(1+\omega^2\tau^2)$~\cite{maxwellmodel} using the least squares with the Marquardt-Levenberg-algorithm adopted in gnuplot (the fitted lines are shown in Fig.~\ref{fig:1} (c)); as a result, we obtained $G = 17.85 \pm 0.17$~Pa and $\tau = 1.23 \pm 0.03$~s. 
We should remark that the flow speed $V$ needs to be lower than the critical one $V_c=h/\tau=0.4$ mm/s to have the outer fluid viscous.
In our experiment, the injection front extends at least one order of magnitude faster than $V_c$, as later shown in Fig.~\ref{fig:5}. Thus, the outer fluid behaves elastically.}

 {\color{black} We here list the relevant physical properties of the samples; viscosity, density, surface tension, and contact angle. The viscosity of inner and outer fluid was 2.08 and 3.03 $\times 10^3$ mPa$\cdot$s. The density of inner and outer fluids was 0.76 and 1.00 g/cm$^3$. We measured the surface tension of inner and outer fluids with the Wilhelmy method (CBVP-A3, Kyowa) to have 26.6 and 31.1 mN/m. We found that the contact angle of the sample under air was almost the same for the top and bottom plates. The contact angle of outer fluids was 30$^\circ$, while the contact angle of inner fluids was not detectable due to the spreading of droplets, indicating both plates are wettable to the inner fluid. However, we also observed that the bottom plate was found to be more wettable from the size of the spread droplet. When the fluorine coating was applied, the contact angles of the inner and outer fluids were approximately 73 and 80$^\circ$.
These measurements were conducted at 22 $\pm 2^\circ$C.}

\section{\label{sec:level3}Results}
In Figure~\ref{fig:2}, we present snapshots of typical fingering patterns observed with $Q=$1.0, 3.0, 6.0, and 9.0 mL/min.
{\color{black} The thick fingers appeared up to 2--3 s, independent of the injection rate $Q$. Then, eventually, the front showed significant splitting, resulting in numerous thin fingers with approximately 1.2--2.2 mm width, as exemplified in the snapshots Fig.~\ref{fig:2}(e). The thin fingers tended to meander and extend not only in radial but also in angular direction from the inlet. }

\begin{figure*}
	\centering
\includegraphics[width=150mm]{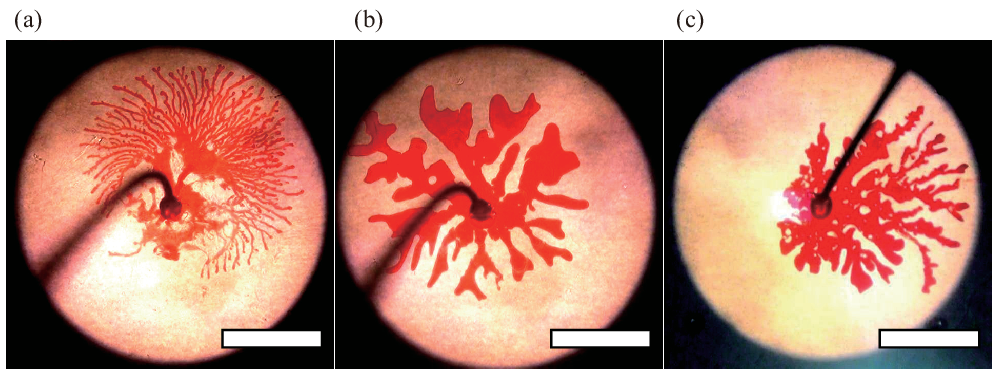}
	\caption{\label{fig:3}
    (a)--(c) Snapshots of the fingering pattern for $Q=$ 9.0 mL/min. at the point the inner fluid reached the edge of the outer fluid. The upper, bottom, and both surfaces were coated with fluorine in (a), (b), and (c), respectively. The fingering pattern was unchanged when the fluorine coat was only applied to the upper plate. Scale bar: 50 mm. {\color{black} Here, we used the outer fluids, the aqueous mixture of CTAB and NaSal mixture with almost identical concentration but with $G$=13 Pa and $\tau$=8 s. }}
\end{figure*}

It's worth noting that such extensive tip splitting and the appearance of thin fingers seem to be related to the wetting of the substrate, which was confirmed with the experiment by altering the surface condition using a fluorocarbon surface modifier. 
{\color{black} The fluorine coating makes the surface less wettable for both the inner and outer fluid; however, the injecting organic fluid seemingly has a larger change. The coating made the surface from almost complete wetting to the contact angle with 73$^\circ$.}
For the comparison, we tried three different situations: the surface modifier was applied only on the upper surface, on the bottom surface, and on both surfaces of the cell (Fig.~\ref{fig:3}(a)--(c)). 
The images shown as Fig.~\ref{fig:3}(b) and (c) clearly showed the effect of the surface modification, whereas the images shown as Fig.~\ref{fig:3}(a) and Fig.~\ref{fig:2}(d) appeared similar, indicating that only the wetting behavior of the bottom surface is related to the thin fingers.
Additionally, the patterns we observed following the surface modification showed thick fingers (Fig.~\ref{fig:3}(b) and (c)), similar to those observed during the initial injection stages, suggesting that the initial thick fingers are insensitive to the surface treatment. Thus, thick fingering must be attributed to the bulk properties of the outer fluid. By contrast, the thin fingers observed later were attributed to the phenomena governed by the wetting properties of the bottom surface.

\begin{figure*}
\centering
\includegraphics[width=150mm]{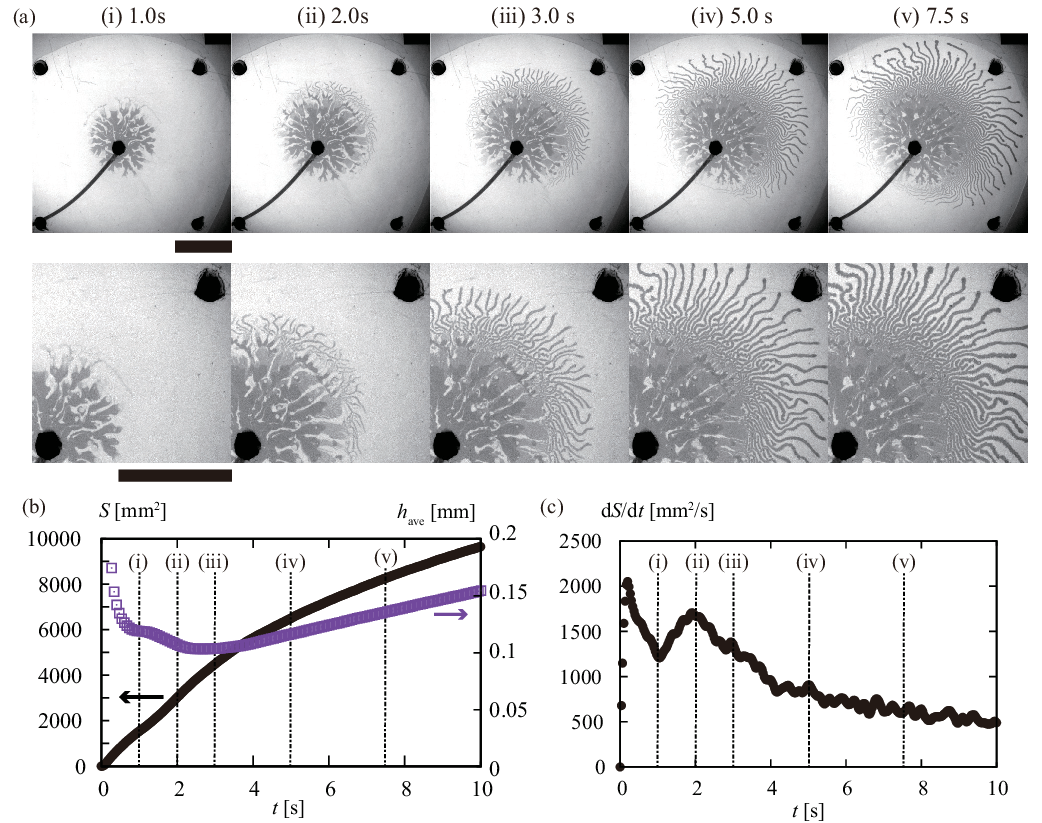}
\caption{\label{fig:4}(a) Snapshots of observed injection pattern with $Q=$ 9.0 mL/min. From 1--2 s (i)--(ii), we observed initially thick fingers, and then after 2 s (ii), thin fingers appeared during the injection. The scale bar corresponds to 50 mm. Time evolution of (b) area $S$ (black) and estimated height of the injected fluid $h_{\mathrm{ave}}$ (purple) for $t$. The appearance of the  (c) time change of area velocity $dS/dt$. Each number corresponds to the same as in panel (a).
}
\end{figure*}

We then examined the characteristics of the fingering dynamics by the image analysis.  In Fig.~\ref{fig:4}(a), we present snapshots of the injection dynamics with $Q=$ 9.0 ml/min.  The transition in the fingering pattern can be observed in Fig.~\ref{fig:4}(a-ii). We obtained the apparent area $S$ of the inner fluid (Fig.~\ref{fig:4}(b) and the estimated average height of injected fluid was obtained from $h_{\mathrm{ave}}=Qt/S$. We also extracted the time evolution $dS/dt$, as shown in Fig.~\ref{fig:4}(c).

Before $t=1$ s corresponding to Fig.~\ref{fig:4}(a-i), $dS/dt$ took an initial peak, where we observed the extension of thick fingers. Considering the existence of time lag in the injection, the observed peak and the subsequent decrease in $dS/dt$ before $t=1$ s reflect the relaxation of excess injection pressure.  Although a stepping motor controlled the constant injection rate at the syringe pump, the rigidity of both the tube and syringe can lead to fluctuation in the pressure of the inner fluid during this initial stage. The initiation of the fluid injection can require relatively high pressure to overcome the Laplace pressure. We also note that the average height of the fluid $h_{\mathrm{ave}}$ took a plateau value of around {\color{black} 0.12 mm}. 

The graph in Fig.~\ref{fig:4}(a-ii) shows another peak in $dS/dt$ around $t=2$ s, which is when thin fingers appear. At this point, the average height $h_{\mathrm{ave}}$ also decreases, reaching a minimum value of around {\color{black} 0.1 mm}. These findings indicate a qualitative difference between the formation of thick and thin fingers. The second peak in $dS/dt$ at $t=2$ s suggests that the wetting of fluid on the substrate can assist in the formation of thin fingers.

\begin{figure*}
	\centering
	\includegraphics[width=150mm]{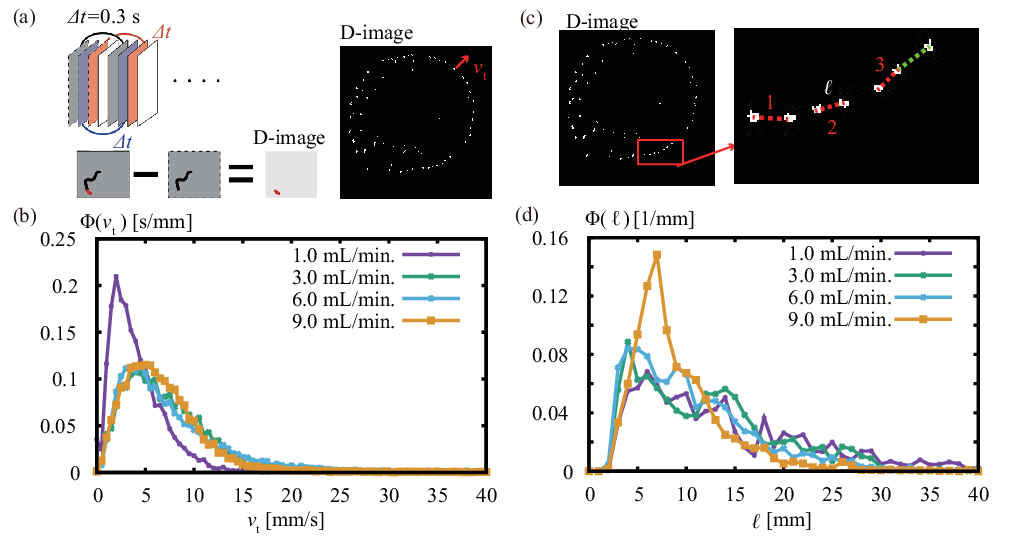}
	\caption{\label{fig:5}(a) Schematic representation of analysis method used {\color{black} with difference image (D-image) where the time interval of {\color{black} $\Delta t$ = 0.3 s} was taken. (b) Normalized distribution of the distance $v_t$ of moving fingers for different values of $Q$. When $Q=$ 3.0--9.0 mL/min, there is a peak at 5 mm/s, whereas the peak is at smaller $v_t$ around 2.5 mm/s in the case of $Q=$ 1.0 mL/min. (c) Schematic representation of analysis method used to find the typical distance $\ell$ between the fingers front. We measured the distance between nearest-neighbor pairs of active fingers connected by colored lines. (d) Normalized distribution of the distance $\ell$ between the active fingers for different values of $Q$. When $Q=$ 1.0--6.0 mL/min, there is a peak in the typical distance around 4 mm, while the peak slightly shifted to 7 mm at $Q=$ 9.0 mL/min}.}
\end{figure*}

To further characterize the pattern of the thin fingers, we focus our attention on the dynamics of the front of the advancing fingers. To do this, we took a temporal difference of the images that were separated by a time interval of {\color{black} $\Delta t$ = 0.3 s} (D-Image). D-Image allowed us to extract snapshots of the advancing fronts from sequential images {\color{black} (Fig.~\ref{fig:5}(a)). For the image analysis, we used the data 3 s after the start of the injection. We detected and traced the spot larger than 0.035 mm$^2$. The spot traced more than 1/3 s was considered as the tip of filaments. With this data, we obtained the speed for each extending filament $v_t$. The normalized distribution of $v_t$, $\phi(v_t)$, for $Q=$ 1.0--9.0 mL/min  is shown in Fig.~\ref{fig:5}(b). Here, $\phi(v_t)$ is normalized such that $\int \phi (v_t) dv_t=1$.} We then used the dame data for detecting the typical distance between neighboring pairs of advancing fingers $\ell$ (Fig.~\ref{fig:5}({\color{black}c})). 

The normalized distribution of distance $\phi(\ell)$ for $Q=$ 1.0--9.0 mL/min  is shown in Fig.~\ref{fig:5}(b).  Here, $\phi(\ell)$ is normalized such that $\int \phi (\ell) d\ell=1$. The distribution of $\phi(\ell)$ is almost the same for {\color{black} $Q=$ 1.0--6.0 mL/min, with a peak at 4 mm. The distribution for $Q=$ 9.0 mL/min also displays a peak, but the position has shifted to a longer distance of 7 mm. }

\section{Discussion}
\begin{figure}[ht]
	\centering
	\includegraphics{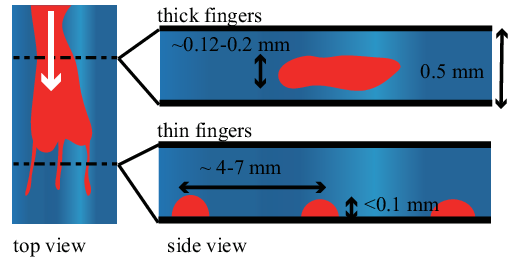}
	\caption{\label{fig:6} Schematic illustration of the thick and thin fingers. Thick fingers appear in the initial stage while fracturing the outer fluid. Thin fingers extend while wetting the bottom surface. 
	}
\end{figure}

In our system, we injected an organic liquid into a cell filled with a viscoelastic fluid, which resulted in the formation of thick and thin fingers. Based on the data we have obtained, we summarize the assumed geometry of these fingers in Fig.~\ref{fig:6}. Here, we suggest that thin fingers should be related to the delamination of outer fluid from the bottom substrate. Initially, the inner fluid causes the bulk fracture of the outer fluid, resulting in thicker fingers. Once the inner fluid reaches the bottom substrate, it starts to delaminate the outer fluid from the bottom substrate, causing thinner fingers. 

{\color{black} Our interpretation is based on the results, where the thin finger was affected by the fluorinated coating of the bottom plate and changed into thick fingers, as shown in Fig.~\ref{fig:3}. This result can be explained by two possibilities: one is the thin finger wet the bottom surface, while the other is the thick one wet the bottom surface. However, based on a previous study on the wetting of elastic film, we can partially justify the above idea: the thin finger is caused by the delamination process. In that study~\cite{Langmuir2003}, they examined the adhesion of an elastic film to a rigid substrate in the air. As the elastic film gradually adhered to the substrate, the main fraction of air that was initially present was expelled from the surface, leaving a fingering pattern with a typical distance of $\lambda_\mathrm{c}=4h$,
where $h$ corresponds to the thickness of the elastic layer. }

Our current study involves injecting oil between the viscoelastic material and the rigid substrate, but it shares similarities with the previous study in that a fluid interacts with a confined elastic layer. In our case, the gap width corresponds to the thickness of the elastic outer fluid, which is 0.5 mm. Therefore, the repeat distance should be $\lambda_\mathrm{c}=2.0$ mm, which is {\color{black} at the same order as the peak observed in Figure~\ref{fig:5}(d) for $Q=$ 1.0--9.0 mL/min.}

{\color{black} We can further support the idea by inspecting the dynamics of the finger extension shown in Fig.~\ref{fig:4}. According to our assumption, the thin finger is attributed to the delamination; the thin finger extension should be enhanced by the wetting of the surface. Indeed, we found an abrupt increase in $dS/dt$ around 2 s (Fig.~\ref{fig:4} (a-ii)), where we found the appearance of thin fingers as seen in Fig.~\ref{fig:4}(c). }

{\color{black} To summarize, the present results show that the injection of an incompressible viscous fluid into a viscoelastic fluid produces fingers. Two types of fingers were observed -- thick and thin, as shown in Fig.~\ref{fig:6}. Thick fingers appear initially and are likely related to bulk fracture formation, which is insensitive to the wetting of the surface. Thin fingers appear later and correspond to the delamination process, where the distance between fingers should be approximately four times the cell depth. Both the delamination process and bulk fracture formation are typical of elastic films. This study reveals that the wettability of a fluid can affect the injection behavior significantly, especially in the case that the injected materials were confined in the other hard substrate. This situation is commonly observed in the industrial and geophysical environment.}

\section{\label{sec:level5}Conclusion}
Our study focuses on exploring the behavior of viscoelastic materials in a Hele-Shaw cell, which can help simulate engineering and geophysical environments. Specifically, we examined the fingering pattern dynamics of a wormlike micellar solution, which was used as an outer viscoelastic fluid in a cell. We used an incompressible viscous liquid, tetradecane stained with oil red, as the inner fluid. The inner fluid was injected from the center of the cell at a fixed rate $Q$, which was chosen to ensure that the outer fluid behaved elastically with slow relaxation, allowing for the injection of fluid irreversibly.

After injecting the inner fluid, we observed a fingering pattern. We observed the different fingering patterns characterized by thick and thin fingers.  It is worth noting that the results of our study differ from previous observations of viscoelastic fingering{\color{black}, which mainly focused on the transition of material viscoelasticity}. In those cases, the transition from viscous to elastic behavior would eventually lead to the formation of bulk fractures, as previously reported in studies~\cite{Mora2010, Lemaire1991}. {\color{black} These fingering do not depend on the surface's wettability. In our study, however, the thin fingers were sensitive to the coating of the bottom plate. The contact angle measurements suggest that the bottom plate was highly wettable to the inner fluid. Based on these observations, we suggest that the delamination of the outer viscoelastic fluid from the bottom surface contributed to the formation of thin fingers. This difference is mainly attributed to using organic inner and aqueous outer fluids with the oleophilic resin as a substrate. Combining these material properties resulted in the inner fluid preferring the bottom surface, which led to the delamination process. This assumption may be confirmed by the three-dimensional analysis of fingers, but technical difficulties make this analysis a future study.}

Our research has shown that modifying the interfacial energy of the substrate surface through chemical treatment can impact fingering behavior. {\color{black} Our sample has a long rheological relaxation time, which made the injection proceed essentially in the elastic regime. However, the wetting of fluid on a surface, which is typically the character of fluid, played an essential role in the fingering process. Thus, our study can be recognized as a type of elastocapillary problem~\cite{elascap}. Our findings are relevant to the injection of fluids into soil even when the injected material behaves as solids. We should note that soil can behave as elastic and is an inhomogeneous surface that is both chemically and geometrically diverse. On this account, further studies could explore how quenched disorder of wetting properties affects fingering dynamics in the elastic environment.}

\begin{acknowledgments}
  The authors would like to thank S. Wagatsuma for his help in constructing the experimental setups at the beginning of this research.
  This work was supported by JSPS KAKENHI Grant Nos. JP16K13866, JP16H06478, 19H05403, 21H00409 and 21H01004. This work was also partially supported by a JSPS Bilateral Joint Research Program between Japan and the Polish Academy of Sciences, ``Spatio-temporal patterns of elements driven by self-generated, geometrically constrained flows,'' and the Cooperative Research of the ``Network Joint Research Center for Materials and Devices'' with Hokkaido University (No. 20181048).
\end{acknowledgments}

\section*{Conflict of Interest Statement}
The authors have no conflicts to disclose.

\section*{Author Contributions}
K.Y. and Y.S. designed research; K.Y., K.O., A.S. and Y.S. performed the experiments; K.Y., K.O., A.S. and Y.S. analyzed data; and K.Y., K.O., A.S. and Y.S. wrote the paper.

\section*{Data Availability Statement}
The data that support the findings of this study are available from the corresponding author upon reasonable request.

\end{document}